\documentclass[letterpaper,prl,twocolumn,showpacs,superscriptaddress]{revtex4}

\usepackage{graphicx}
\usepackage{hyperref}
\usepackage{bm}
\usepackage{amsmath}

\renewcommand{\section}[1]{{\par\it #1.---}}

\begin{document}

\title
{Gauging a quantum heat bath with dissipative Landau-Zener transitions}

\author{Martijn Wubs}
\affiliation{Institut f\"{u}r Physik, Universit\"{a}t Augsburg,
Universit\"{a}tsstra{\ss}e 1, D-86135 Augsburg, Germany}

\author{Keiji Saito}
\affiliation{Department of Physics, Graduate School of Science,
University of Tokyo, Tokyo 113-0033, Japan}

\author{Sigmund Kohler}
\affiliation{Institut f\"{u}r Physik, Universit\"{a}t Augsburg,
Universit\"{a}tsstra{\ss}e 1, D-86135 Augsburg, Germany}

\author{Peter H\"{a}nggi}
\affiliation{Institut f\"{u}r Physik, Universit\"{a}t Augsburg,
Universit\"{a}tsstra{\ss}e 1, D-86135 Augsburg, Germany}

\author{Yosuke Kayanuma}
\affiliation{Department of Mathematical Science, Graduate School of
Engineering, Osaka Prefecture University, Sakai 599-8531, Japan}

\date{\today}

\begin{abstract}
We calculate the {\em exact} Landau-Zener transitions probabilities for a qubit with arbitrary linear coupling to a bath at zero temperature. 
The final quantum state exhibits a peculiar entanglement between the qubit and the bath. 
In the special case of a diagonal coupling, the bath does not influence the transition probability, whatever the speed of the Landau-Zener sweep. It is proposed to use Landau-Zener transitions to determine both the reorganization energy and the integrated spectral density of the bath. Possible applications include circuit QED and molecular nanomagnets.

\end{abstract}


\maketitle
%
Quite a number of quantum two-state systems are presently tested as candidate qubits, the units of quantum information. Good qubits are well isolated from their environment, but easy to manipulate. This somewhat conflicting requirement has spurred renewed interest in the dynamics of qubits coupled to an environment or heat bath \cite{Leggett1987a,Weiss1999a}.  Qubits can be seen as bath detectors. For example, static qubits probe via their decay rates the bath spectral density at their transition frequencies. In solid-state environments these rates can be strongly frequency- and sample-dependent. We discuss an in this respect superior  `bath detection mode' of the qubit. 

 One way of changing the state of a two-level system involves the forced crossing of its diabatic energies. For constant level-crossing speed this is  
 known as the Landau-Zener (LZ) problem \cite{Landau1932a,Zener1932a,Stueckelberg1932a}, which for a two-level system can be solved exactly 
 \cite{Landau1932a,Zener1932a,Stueckelberg1932a,Kayanuma1984a}. This is no longer the case when taking its environment into account 
 \cite{Gefen1987a,Ao1989a,Ao1991a,Kayanuma1998a}  that may cause thermal excitation and quantum tunneling. 
  In the low-temperature tunneling regime, analytical estimates for transition probabilities exist only for very fast and very slow sweeps,
  and the literature is not unanimous about the latter limit \cite{Gefen1987a,Ao1989a,Ao1991a,Kayanuma1998a,Grifoni1998a}. 

In another line of research, incited by the paper \cite{Brundobler1993a}, it was recently proven  that LZ transition probabilities can be calculated exactly  for some many-level systems as well, although only for some initial states \cite{Shytov2004a,Sinitsyn2004a,Volkov2005a,Saito2006b}. In this Letter, we extend the analysis to quantum dissipative systems and study LZ transitions in spin-boson problems in a new way. First we calculate zero-temperature LZ transition probabilities exactly. Second, by making use of this exact dependence on the bath parameters, we propose to gauge the dissipative environment of a qubit by performing LZ sweeps. One advantage of this bath detection mode of the qubit is that effects of spiky bath spectral densities are averaged out in every sweep.

%
\section{Driven spin-boson model}
Consider LZ transitions in a qubit coupled to a bath of $N$ quantum harmonic oscillators, as described by the Hamiltonian 
\begin{equation}\label{e.Ham1}
H(t)  = \frac{v t}{2} \bm{\sigma}_{z} + \frac{\Delta}{2}
\bm{\sigma}_{x} +
\sum_{j=1}^{N} \hbar \Omega_{j} b_{j}^{\dag}b_{j} + H_{\mathrm{QO}}, 
\end{equation}
with the qubit-oscillator coupling 
\begin{equation}\label{e.HQO}
H_{\mathrm{QO}}  =  \sum_{j=1}^{N}\frac{\gamma_{j}}{2}\left[\cos
\theta_{j}\bm{\sigma}_{z}+ \sin
\theta_{j}\bm{\sigma}_{x}\right](b_{j}+b_{j}^{\dag}).
\end{equation}
The energy difference between the diabatic qubit states changes linearly in time as $v t$ (with level-crossing speed $v>0)$ and their intrinsic interaction amplitude is $\Delta$. The ${\bm \sigma}_{x,z}$ are Pauli operators. The first two terms of (\ref{e.Ham1}) define the standard Landau-Zener problem for an isolated two-level system.  The $N$ harmonic oscillators in (\ref{e.Ham1}) can have different frequencies $\Omega_{j}>0$, qubit-oscillator couplings $\gamma_{j}$, and interaction angles $\theta_{j}$. 
The oscillators affect the qubit  (i) by changing its energies via the diagonal coupling $\propto \gamma_{j}\cos \theta_{j}\bm{\sigma}_{z}$, and (ii) by inducing transitions between its levels via the transverse coupling $\propto \gamma_{j}\sin \theta_{j}\bm{\sigma}_{x}$. 

In order to  calculate the probability that the qubit state flips due to the LZ sweep, we will work in an interaction picture and split the Hamiltonian~(\ref{e.Ham1}) into  an interaction $V = {\bm \sigma}_{x}[\Delta/2+ \sum_{j=1}^{N}
(\gamma_{j}/2)\sin
\theta_{j}(b_{j}+b_{j}^{\dag})]$ involving bit flips as described by  ${\bm \sigma}_{x}=|{\uparrow}\rangle\langle{ \downarrow}| + |{\downarrow}\rangle\langle{\uparrow}|$, and the (bit-flip-)free Hamiltonian $H_{0}(t)$. A polaron transformation \cite{Weiss1999a} diagonalizes $H_{0}(t)$ in terms of qubit-state dependent shifted oscillators, giving
$H_{0}(t)$ the form $\sum_{s = \uparrow,\downarrow} |{s}\rangle\langle{s}|
\bigl\{ \pm \frac{vt }{2} + \sum_{j=1}^{N} \hbar \Omega_{j}
b_{j\pm}^{\dag}b_{j\pm}-E\bigl\}$,
with creation and annihilation operators for the shifted oscillators $b_{j\pm}^{(\dag)}  \equiv   b_{j}^{(\dag)} \pm Q_{j}/2$ with 
$Q_{j} \equiv (\gamma_{j}/\hbar \Omega_{j})\cos\theta_{j}$; the  $+$ corresponds to the qubit state $|{\uparrow}\rangle$ and the $-$ to $|{\downarrow}\rangle$. With bath oscillators shifting in this way, the reorganization energy~\cite{Weiss1999a} gained by the system has the same value  $E=\sum_{j=1}^{N} (\gamma_{j}^{2}/4 \hbar \Omega_{j})\cos^{2}\theta_{j}$ for both qubit states.
Eigenstates of the shifted oscillators are labelled as $|{{\bf n}_{\pm}}\rangle$, where the $N$ components $n_{j}$  of the vector ${\bf n}$ are single-oscillator excitation numbers. 
The free time-evolution operator $U_{0}(t)$ can be written as
$\sum_{s=\uparrow,\downarrow}|s\rangle\langle s|e^{\mp \mathrm{i} v
t^{2}/4\hbar}\sum_{{\bf n}}|{\bf n}_{\pm}\rangle\langle {\bf
n}_{\pm}|\ e^{- \mathrm{i} {\bf n}\cdot{\bm \Omega} t + \mathrm{i}  E t/\hbar}$,
where the inner product of the vectors ${\bf n}$ and ${\bm \Omega}=(\Omega_{1},\Omega_{2},\ldots,\Omega_{N})$ shows up. 
We next define the interaction-picture Hamiltonian as $\tilde H(t)  \equiv   U_{0}^{\dag}(t)\,V\,U_{0}(t)$. In order to bring $\tilde H(t)$ into a useful form, we write the oscillator operators  $(b_{j}+b_{j}^{\dag})$ in $V$ as $(b_{j\pm}+b_{j\pm}^{\dag}\mp Q_{j})$. We choose to associate the  ``$+$''-oscillators with the term $|{\downarrow}\rangle\langle{\uparrow}|$ of  ${\bm \sigma}_{x}$, and the ``$-$''-operators with the other term.  We then write $b_{j\pm} = \sum_{\bf n}\sqrt{n_{j}}|{({\bf n}-{\bm 1}_{j})_{\pm}}\rangle\langle{ {\bf n}_{\pm}}|$, where  ${\bm 1}_{j}$ is the unit vector with $j{\mathrm{th}}$ component equal to $1$, and likewise $b_{j\pm}^{\dag} = \sum_{\bf n}\sqrt{n_{j}+1}|{({\bf n}+{\bm 1}_{j})_{\pm}}\rangle\langle{ {\bf n}_{\pm}}|$.  In this form,  the `bra's' of shifted oscillator states in $V$ correspond to the `bra's' of the qubit, while for the `kets' this must still be arranged  by using the completeness relation 
$|{{\bf m}_{\pm}}\rangle = \sum_{{\bf n}} |{{\bf n}_{\mp}}\rangle \langle
{{\bf n}_{\mp}}|{{\bf m}_{\pm}}\rangle$. 
The interaction-picture Hamiltonian then becomes
\begin{eqnarray}\label{tildeHt}
\tilde H(t) & = & \frac{1}{2}\sum_{{\bf m},{\bf n}}e^{\mathrm{i} ({\bf m}-{\bf n})\cdot{\bm \Omega} t}\bigl\{ W_{{\bf m}{\bf n}}^{+}
|{\uparrow}\rangle\langle{\downarrow}|e^{ \mathrm{i} v t^{2}/2\hbar}\otimes
|{{\bf m}_{+}}\rangle\langle
{{\bf n}_{-}}| \nonumber \\
&& + W_{{\bf m}{\bf n}}^{-}|{\downarrow}\rangle\langle{\uparrow}|e^{- \mathrm{i} v
t^{2}/2\hbar}\otimes |{{\bf m}_{-}}\rangle\langle{ {\bf n}_{+}}|\bigl\},
\end{eqnarray}  
involving the two infinite-dimensional matrices $W^{\pm}$ with
 \begin{widetext}
\begin{equation}\label{Wpm} 
W_{{\bf
m}{\bf n}}^{\pm} =   \bigg(\Delta \pm
\sum_{j=1}^{N} \gamma_{j}Q_{j}\sin\theta_{j}\bigg)\langle
{{\bf m}_{\pm}}|{{\bf n}_{\mp}}\rangle 
 + \sum_{j=1}^{N}\gamma_{j}\sin\theta_{j}
\bigl\{\sqrt{n_{j}}\langle{{\bf m}_{\pm}}|{({\bf n}-{\bm 1}_{j})_{\mp}}\rangle  
+ 
\sqrt{n_{j}+1}\langle{{\bf m}_{\pm}}|{({\bf n}+{\bm 1}_{j})_{\mp}}\rangle \bigl\}. 
\end{equation}

\section{Transition probability}
We focus on the situation that  at time $t=-\infty$ the system starts in its ground
state $|{\psi(-\infty)}\rangle = |{\uparrow}\rangle\,|{{\bm 0}_{+}}\rangle$.
We are now interested in the survival probability $P_{{\uparrow\rightarrow\uparrow}}(\infty)$ of the initial state
$|{\uparrow}\rangle$ of the qubit. $P_{{\uparrow\rightarrow\uparrow}}(\infty)$ equals the square of the norm of the projected final oscillator state
$\langle{\uparrow}|{\tilde\psi(\infty)}\rangle$, which can be written as $\langle{\uparrow}| \tilde
U(\infty,-\infty)|{\uparrow}\rangle\,|{{\bm 0}_{+}}\rangle$, where  $\tilde U(t_{2},t_{1}) = \overleftarrow {T}\,\exp[-({\mathrm
i}/\hbar)\int_{t_{1}}^{t_{2}}\mbox{d} \tau\,\tilde H(\tau)]$ is the
time evolution operator in the interaction picture. In a time-ordered expansion of $\tilde U(\infty,-\infty)$, only the even powers of $\tilde H(\tau)$ will contribute to $P_{{\uparrow\rightarrow\uparrow}}(\infty)$.
As is well known, the $(2k){\mathrm{th}}$-order term in the expansion involves a  $(2k)$-fold time integral with variables $t_{2k}>t_{2k-1}>\ldots>t_{2}>t_{1}$ in the interval ($-\infty,\infty$).  It is advantageous to make the variable transformation~\cite{Kayanuma1984a} $x_{q}=t_{1} + \sum_{\ell = 1}^{q-1}(t_{2\ell +1}-t_{2\ell})$ for $2
\le q\le k$, $x_{1}=t_{1}$, and $y_{q} = t_{q}-t_{2q-1}$. 
We label the $N$-oscillator state after $\ell$ interactions as ${\bf n}^{(\ell)}$.
For brevity, we define the frequencies $w_{\ell} = ({\bf n}^{(\ell)}-{\bf n}^{(\ell-1)})\cdot {\bm \Omega}$ and $w_{1}={\bf n}^{(1)}\cdot {\bm \Omega}$. 
The perturbation series for $\langle{\uparrow}|{\tilde\psi(\infty)}\rangle$ then becomes
\begin{eqnarray}\label{e.integralsxy}
& & \sum_{k=0}^{\infty}\frac{1}{(2 \mathrm{i}\hbar)^{2k}}\int_{-\infty}^{\infty}\mbox{d}
x_{1}\int_{x_{1}}^{\infty}\mbox{d} x_{2}\ldots
\int_{x_{k-1}}^{\infty}\mbox{d} x_{k} \int_{0}^{\infty}\mbox{d}
y_{1}
\ldots \int_{0}^{\infty}\mbox{d} y_{k} 
\sum_{{\bf n}^{(2k)},
\ldots,
{\bf n}^{(1)}} 
W_{{\bf n}^{(2k)}{\bf n}^{(2k-1)}}^{\rm +} W_{{\bf
n}^{(2k-1)}{\bf n}^{(2k-2)}}^{\rm -}\ldots W_{{\bf
n}^{(2)}{\bf n}^{(1)}}^{\rm +}
W_{{\bf n}^{(1)}{\bm 0}}^{\rm -}  \nonumber \\
&& \times 
\exp\left[\frac{\mathrm{i}v}{\hbar}\sum_{l=1}^{k}
\left( 
x_{\ell}y_{\ell}
 +\frac{\hbar}{v}(w_{2\ell}+w_{2\ell -1})x_{\ell}
  + \frac{1}{2}y_{\ell}^{2}
+y_{\ell}\sum_{q=1}^{\ell-1}y_{q}
 \right)
 + \mathrm{i}\sum_{\ell = 1}^{k}
 \left(
   w_{2\ell} y_{\ell}
 +(w_{2\ell}+w_{2\ell -1})\sum_{q=1}^{\ell-1}y_{q}
 \right) \right] |{{\bf n}^{(2k)}_{+}}\rangle. 
 \end{eqnarray}
\end{widetext}
It would be convenient to symmetrize the $x_{\ell}$-integrals at this point, but in general the integrand is not symmetric under permutation of the $x_{\ell}$. When transforming the $x_{\ell}$ to new variables $s_{1}=x_{1}$ and $s_{\ell}=x_{\ell}-x_{\ell-1}$ for $\ell =2,3,\ldots,k$, one finds that the  $\int_{-\infty}^{\infty}\mbox{d}s_{1}$-integral yields the delta-function $(2\pi\hbar/v)\delta(\sum_{\ell = 1}^{k} y_{\ell} + \frac{v}{\hbar}(w_{2\ell}+w_{2\ell-1})\,)$. Now, since the initial state is $|{\uparrow}\rangle\,|{{\bm 0}_{+}}\rangle$, the sum over the $w_{\ell}$ in this delta function equals ${\bf n}^{(2k)}\cdot{\bm  \Omega}$, which evidently is $\ge 0$. Likewise, the variables $y_{\ell}$ are all positive or zero. Therefore, the delta-function can only ``click'' in the subspace $y_{1}=y_{2}=\ldots =y_{2k}=0$ and will do so only if the vector ${\bf n}^{(2k)}$ vanishes. The physical meaning of the latter statement is discussed below. 
Performing the other $s_{\ell}$-integrals is cumbersome, so we quickly return to the integrals in Eq.~(\ref{e.integralsxy}), but this time armed with the knowledge that only the subspace  $y_{1}=\ldots =y_{2k}=0$ contributes. Since within this subspace the integrand {\em is} symmetric in the variables $x_{\ell}$, it is correct to symmetrize the $x_{\ell}$-integrals in Eq.~(\ref{e.integralsxy}), i.e. we can replace them by $(1/k!)\int_{-\infty}^{\infty}\mbox{d}
x_{1}\ldots
\int_{-\infty}^{\infty}\mbox{d} x_{k}$. After performing these standard integrals,
the $y_{\ell}$-integrals can be evaluated as well. 
For example, the $y_{1}$-integral 
$\int_{0}^{\infty}\mbox{d}y_{1}\,\delta(y_{1}+ \hbar w_{2}/v + \hbar w_{1}/v)$ vanishes unless  
$(w_{2}+w_{1})={\bf n}^{(2)}\cdot {\bm \Omega}$ vanishes, in which case it is indeed the ``subspace'' defined by $y_{1}=0$ that makes this integral equal to $\frac{1}{2}$. 

From the time integrals we find the following {\em selection rule}\,:
 when starting in the ground state $|{\uparrow}\rangle\,|{{\bm 0}_{+}}\rangle$, the only $(2k){\mathrm{th}}$-order processes 
 contributing to the survival probability $P_{{\uparrow\rightarrow \uparrow}}(\infty)$ are those with 
${\bf n}^{(2)}={\bf n}^{(4)}=\ldots={\bf n}^{(2k)}={\bm 0}$. Hence the oscillators will end up in their initial state $|{{\bm 0}_{+}}\rangle$ in case the qubit ends up in $|{\uparrow}\rangle$. This striking result agrees with  the so-called no-go theorem \cite{Brundobler1993a,Shytov2004a,Sinitsyn2004a,Volkov2005a}, which we extended here to spin-boson problems. The time integrals do not forbid occupation  of the states $|{\uparrow}\rangle|{{\bf n}_{+}\ne {\bm 0}_{+}}\rangle$ at intermediate times, nor do they restrict the intermediate oscillator states  $|{\bf n}_{-}^{(2\ell-1)}\rangle$, but further restrictions may originate from vanishing matrix elements $W_{\bf mn}^{\pm}$, see Eq.~(\ref{Wpm}). Only when the qubit ends up in $|{\downarrow}\rangle$ can the LZ driving dissipate energy into the bath. Hence qubit and bath end up entangled.

In line with the selection rule, we find that $\langle{\uparrow}|{\tilde \psi(\infty)}\rangle$ simplifies into 
$ \exp(-\pi W^{2}/4\hbar v)\,|{\bm
0}_{+}\rangle$, where the parameter $W^{2} \equiv \sum_{{\bf n}}W_{{\bm 0}{\bf n}}^{\rm +} W_{{\bf n}{\bm
0}}^{\rm -}$ is still to be determined  by using Eq.~(\ref{Wpm}). 
We finally obtain our central result: The exact Landau-Zener transition probability for a qubit arbitrarily coupled to an oscillator bath at $T=0$ is
\begin{equation}\label{transitionsexact}
P_{{\uparrow \rightarrow \downarrow}}(\infty)=1- P_{{\uparrow \rightarrow \uparrow}}(\infty) =1- e^{-\pi W^{2}/(2\hbar
v)}, 
\end{equation}
where the parameter $W^{2}$  is given by
\begin{equation}\label{W2final}
W^{2}  =  \bigg( \Delta -
\sum_{j=1}^{N}\frac{\sin\theta_{j}\cos\theta_{j}\gamma_{j}^{2}}{\hbar
\Omega_{j}}\bigg)^2 \,+ \,\sum_{j=1}^{N}
\sin^{2}\theta_{j}\,\gamma_{j}^{2}. 
\end{equation}
By introducing the spectral density $J(\omega) = \sum_{j=1}^{N} (2\gamma_{j}/\hbar)^{2}\delta(\omega - \Omega_{j})$ and assuming that oscillators with equal frequencies $\omega$ have equal coupling angles $\theta(\omega)$, Eq.~(\ref{W2final}) becomes
\begin{eqnarray}\label{W2finalcontinuum}
W^{2} & = & \Big( \Delta -
\frac{\hbar}{4\pi}\int_{0}^{\infty}\mbox{d}\omega\,\sin[\theta(\omega)]\cos[\theta(\omega)]J(\omega)/\omega
\Big)^2 \nonumber \\
&  & +\frac{\hbar^{2}}{4\pi}\int_{0}^{\infty}\mbox{d}\omega\,\sin^{2}[\theta(\omega)]J(\omega).
\end{eqnarray}
 No specific form of  $J(\omega)$ or  $\theta(\omega)$ is presupposed, but for the limit $N \rightarrow \infty$ describing a continuum of oscillators, 
the two integrals in Eq.~(\ref{W2finalcontinuum}) have to be finite.

%
\section{Transverse coupling ($\theta_{j}=\pi/2$)}
We first focus on purely transverse bath coupling and obtain $W^{2} =  \Delta^{2}+S$, where 
$S=\sum_{j=1}^{N}\gamma_{j}^{2}=(\hbar^{2}/(4\pi))\int_{0}^{\infty}\mbox{d}\omega J(\omega)$
is the integrated spectral density. The $j{\mathrm{th}}$ oscillator reduces the final survival probability by a factor $\exp[-\pi \gamma_{j}^{2}/(2\hbar v)]$, independent of the other oscillators or of the value of $\Omega_{j}$.  For $\Delta=0$, the bath ends in $|{{\bm 0}_{+}}\rangle$ when the qubit ends in $|{\uparrow}\rangle$, whereas the bath contains an odd number of bosons when the qubit ends in 
$|{\uparrow}\rangle$. Qubit and bath therefore end up fully entangled,
in the sense that the qubit coherence vanishes after tracing out the bath. 
Nonzero final qubit coherence requires  $\Delta\ne 0$.   Notice also the interesting phenomenon of {\em bath-assisted adiabatic following}: For large bath coupling, i.e. $S \gg 2\hbar v/\pi$, the qubit
ends up in the initially unpopulated state,
even in the absence of an intrinsic interaction $\Delta$. 
%

\section{Diagonal coupling ($\theta_{j}=0$)}
In solid-state environments, decoherence often occurs much faster than the bath-induced relaxation so that the latter is neglected. Hence the ``standard model'' for LZ transitions in dissipative environments~\cite{Gefen1987a,Ao1989a,Ao1991a,Kayanuma1998a} is the single qubit diagonally coupled to an oscillator bath, which is obtained from our Hamiltonian~(\ref{e.Ham1}) by setting all $\theta_{j}=0$.  By doing the same in Eq.~(\ref{W2final}), we find that for a qubit diagonally coupled to a bath at $T=0$, the Landau-Zener transition probability $P_{{\uparrow \rightarrow \downarrow}}(\infty)=1- \exp[-\pi \Delta^{2}/(2\hbar v)]$, which is the well-known  result for an {\em isolated} qubit! Thus, although the bath coupling does not commute with the qubit Hamiltonian, there is no bath dependence in the transition probability, neither via qubit-bath couplings $\gamma_{j}$ nor via the oscillator frequencies $\Omega_{j}$, no matter how fast the LZ sweep is performed. We discuss this below.

\section{Gauging  quantum dissipation}
Next assume that all oscillators couple to the qubit with the same angle $\theta_{j}=\theta$. This may follow from microscopic considerations or  may be engineered.
For example, when all oscillators couple diagonally to the qubit in a basis described by Pauli matrices ${\bm \tau}_{x,y,z}$, then an experiment where the qubit is driven via $(vt/2){\bm \sigma}_{z}=  (vt/2)(\cos\theta {\bm \tau}_{z}-\sin\theta{\bm \tau}_{x})$ while $(\Delta/2){\bm \sigma}_{x}= (\Delta/2)(\sin\theta {\bm \tau}_{z}+\cos\theta{\bm \tau}_{x})$ is kept constant, is described by our Hamiltonian~(\ref{e.Ham1}) with all angles  $\theta_{j}=\theta$. The reorganization energy then becomes $ E = E_{0}\cos^{2}\theta $, where 
$E_{0} = \sum_{j=1}^{N} \gamma_{j}^{2}/(4 \hbar \Omega_{j})=(\hbar/4\pi)\int_{0}^{\infty}\mbox{d}\omega\,J(\omega)/\omega$ occurs for diagonal coupling. From Eq.~(\ref{W2final}) we deduce
\begin{equation}\label{W2finaltheta}
W^{2} = \left( \Delta -
  E_{0}\sin\theta\cos\theta \right)^2 \,+ \, S \sin^{2}\theta. 
\end{equation}
Unlike for transverse coupling, the oscillators do not affect $P_{{\uparrow \rightarrow \uparrow}}(\infty)$ independently. Relaxation $\propto \sin\theta$, dephasing $\propto \cos\theta$, and intrinsic interaction act together. In the continuum limit $N\rightarrow \infty$, Eq.~(\ref{W2finaltheta}) holds for spectral densities $J(\omega)$ that give rise to finite reorganization energies $E_{0}$ and finite integrated spectral densities $S$.

\begin{figure}[t]
\centerline{\includegraphics{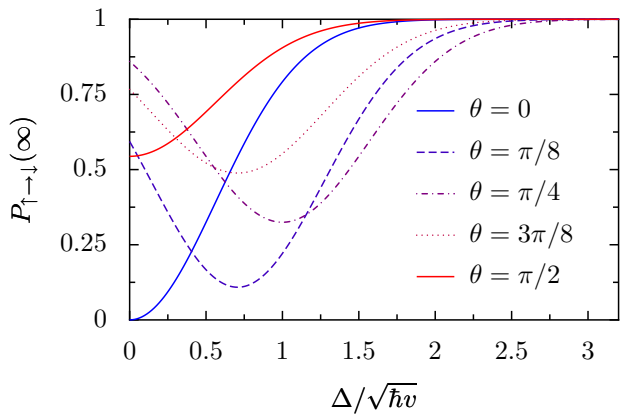}}
\caption{(Color online)
Final transition probability $P_{\uparrow \rightarrow \downarrow}(\infty)$ as a  function of intrinsic interaction $\Delta$, for several values of the coupling angle $\theta$. Parameters: $E_{0}=2.0\sqrt{\hbar v}$ and $S = 0.5 \hbar v$. 
}
\label{fig:gauging}
\end{figure}

As a main application of our calculations, we propose to gauge the dissipative environment of a qubit via LZ transitions. By gauging we mean in short: the measurement of $E_{0}$ and $S$ and the subsequent parameter fixing of appropriate model spectral densities. More in detail, we propose to perform LZ sweeps and to determine transition probabilities for several fixed values of the tunable intrinsic interaction $\Delta$.  Notice that  $W^{2}(\Delta)$ as a function of $\Delta$ is a parabola in Eq.~(\ref{W2finaltheta}). Figure \ref{fig:gauging} 
depicts corresponding final transition probabilities for several coupling angles. Diagonal ($\theta_{j} = 0$) and transverse ($\theta_{j}=\pi/2$) coupling are limiting cases which have in common that $W^{2}(\Delta)$ and hence  $P_{{\uparrow \rightarrow \downarrow}}(\infty)$ 
assume their minima for $\Delta = 0$. Neither limiting case is suitable for gauging $E_{0}$. Since for intermediate cases $0 < \theta < \pi/2$ we find a minimal transition probability for a {\em nonzero} internal interaction $\Delta_{\mathrm{min}} = E_{0}\sin\theta\cos\theta$, one can determine the reorganization energy $E_{0}$
by measuring $\Delta_{\mathrm{min}}$.  Moreover, the integrated spectral density $S$ can then be identified as $S=  W^{2}(\Delta_{\mathrm{min}})/\sin^{2}\theta$.  As a consistency test, one can check   whether $W^{2}(\Delta=0) = \sin^{2}\theta(S+ E_{0}^{2}\cos^{2}\theta)$ holds. Consistency can be further tested by changing  the basis in which the LZ sweep is performed: the values  of $E_{0}$ and $S$ should come out independent of the angle $\theta$. If not, then $\theta(\omega)$ is not  constant and  Eq.~(\ref{W2finalcontinuum}) applies.

With the values of $E_{0}$ and $S$ thus determined, one can fix parameters in appropriate model spectral densities. For example, suppose $J(\omega)$ has the known form $J(\omega)=\alpha \omega^{s} e^{-\omega/\omega_{c}}$ with a power $s>0$ given by the physical nature of the environment \cite{Weiss1999a}, but with unkown strength $\alpha$ and cutoff frequency $\omega_{c}$. ($s=1$ is the Ohmic case and for $s\le 0$,  $E_{0}$ would diverge.) Then 
$E_{0} = \alpha\hbar\omega_{c}^{s}\Gamma(s)/(4\pi)$, in terms of the gamma function, and 
$S= \alpha\hbar^{2}\omega_{c}^{s+1} \Gamma(s+1)/(4\pi)$. Hence LZ gauging  fixes  $\hbar\omega_{c}=S/(E_{0}s)$ and $\alpha$.  Other  spectral densities  may depend on more parameters. LZ gauging fixes two of them.

The efficient use of our gauging scenario requires the following: For $P_{{\uparrow \rightarrow \downarrow}}(\infty)$ to change considerably, one chooses $v$ such that $(\pi S/2\hbar v)\sin^{2}\theta\simeq 1$ and varies $\Delta$ on the scale of $\sqrt{S}\sin\theta$. Then $E_{0}$ can be measured accurately if $E_{0}\cos\theta$ is not much smaller than $\sqrt{S}$, or $\sqrt{\alpha/4\pi}$ not much smaller than unity for an Ohmic bath. So  $P_{{\uparrow \rightarrow \downarrow}}(\infty)$ is robust under dephasing, even if dephasing is much faster than relaxation ($\theta \ll 1$), unless the qubit-bath coupling is strong. This robustness has found its use in experiments on molecular nanomagnets \cite{Wernsdorfer1999a,Leuenberger2000a}.
 
\section{Discussion and conclusions}
We presented in Eqs.~(\ref{transitionsexact}) and (\ref{W2final}) exact Landau-Zener transition probabilities  for a qubit with arbitrary linear  coupling to a zero-temperature bath. 
We found that qubit and bath end up entangled. Our results apply to experiments where the initial and final qubit energies are off-resonant with relevant bath frequencies, for example in circuit QED \cite{Chiorescu2004a,Wallraff2004a},  where qubit energies can be varied over a broad range and spectral densities are peaked \cite{Tian2002a,VanderWal2003a}. Indeed, our predictions for transverse coupling generalize our detailed study of LZ sweeps in circuit QED \cite{Saito2006b} to more  realistic situations where  peaked spectral densities have nonzero widths.  Other applications include tunable atoms in optical cavities and in photonic crystals.

For diagonal coupling we find that the transition probability does not depend on the bath at all. Our exact result settles a long-standing discussion \cite{Gefen1987a,Ao1989a,Ao1991a,Kayanuma1998a,Grifoni1998a}, at least for zero temperature. It corroborates and interpolates between what Ao and Rammer \cite{Ao1989a,Ao1991a} found  for the fast-passage limit $\Delta^{2}/\hbar v \ll 1$ and for the opposite adiabatic limit $\Delta^{2}/\hbar v \ll 1$, which both have been confirmed numerically \cite{Kayanuma1998a}. For fast LZ sweeps,   the bath clearly has no time to affect the transition, but the absence of any bath influence also for slower sweeps is a  highly nontrivial property of this standard model.

A qubit undergoing a LZ sweep measures `global' properties of the bath, namely the frequency-integrated spectral density, and for strong coupling the reorganization energy. Sample-dependent spikes in spectral densities are therefore averaged out, even in a single sweep, so that  model parameters can be determined. We therefore propose to employ LZ transitions for a valuable gauging of the dissipative environment of a tunable qubit.  

\begin{acknowledgments}
This work has been supported by the DFG through SFB\,484 and SFB\,631, and by a
Grant-in-Aid from the Ministry of Education, Sciences, Sports, Culture and Technology of
Japan (No.\ 18540323).
\end{acknowledgments}

\bibliographystyle{prsty}

\end{document}